\documentstyle[twocolumn,aps,epsfig]{revtex}

    \def\z{\noindent}  
 
    \def\sqr#1#2{{\vcenter{\vbox{\hrule height .#2pt
                             \hbox{\vrule width .#2pt height#1pt \kern#1pt
                                   \vrule width .#2pt}
                             \hrule height .#2pt}}}}

    \def\ZZ{\bf{Z}}

\begin{document}
\wideabs{ \title{Ionization of a Model Atom: Exact Results and Connection
with Experiment}

\author{O. Costin, J. L. Lebowitz\footnote{Also Department of Physics}, and
A. Rokhlenko } \address{Department of Mathematics\\ Rutgers University
\\ Piscataway, NJ 08854-8019}

\maketitle
\begin{abstract}
 We prove that a model atom having one bound state will be fully ionized
  by a time periodic potential of arbitrary strength $r$ and frequency
  $\omega$.  The survival probability is for small $r$ given by
  $e^{-\Gamma t}$ for times of order $\Gamma^{-1}$ $\sim r^{-2n}$, where
  $n$ is the number of ``photons'' required for ionization, with
  enhanced stability at resonances.  For late times the decay is like
  $t^{-3}$.  Results are  for a 1d system with a delta
  function potential of strength $-g(1 + \eta (t))$ but comparison
  with experiments on the microwave ionization of excited hydrogen atoms
  and with recent analytical work indicate that many features are
  universal.

\smallskip

\z PACS: 32.80.Rm, 03.65.Db, 32.80.Wr.
\end{abstract}

\centerline{*******}}

\narrowtext

Transitions between bound and free states of a system are of great
importance in many areas of science \cite{[1]} and ``much of the
practical business of quantum mechanics is calculating exponential decay
rates'' \cite{[2]}.  There are, however, still many unresolved questions
when one goes beyond perturbation theory \cite{[1]}--\cite{[6]}.
Unfortunately, approaches going beyond perturbation theory such as
Floquet theory, semi-classical analysis and numerical solution of the
time dependent Schr\"odinger equation are both complicated and also
involve, when calculating transitions to the continuum, uncontrolled
approximations \cite{[1]}--\cite{[5a]}.  It is only recently that some general results
going beyond perturbation theory have been rigorously established for
models with spatial structure \cite{[6]}.  We still don't know, however,
many basic facts about the ionization process, e.g. the conditions for a
time dependent external field to fully dissociate a molecule or ionize
an atom, much less the ionization probability as a function of time and
of the form of such a field \cite{[7]}. Granted that the problem is
intrinsically complicated it would be very valuable to have some simple
solvable models which contain the spatial structure of the bound state
and the continuum and can thus serve as a guide to the essential
features of the process.

In this note we describe new exact results relating to ionization of a
very simple model atom by an oscillating field (potential) of arbitrary
strength and frequency.  While our results hold for arbitrary strength
perturbations, the predictions are particularly explicit and sharp in
the case where the strength of the oscillating field is small relative
to the binding potential---a situation commonly encountered in practice.
Going beyond perturbation theory we rigorously prove the existence of a
well defined exponential decay regime which is followed, for late times
when the survival probability is already very low, by a power law decay.
This is true no matter how small the frequency. The times required for
ionization are however very dependent on the perturbing frequency.  For
a harmonic perturbation with frequency $\omega$ the logarithm of the
ionization time grows like $r^{-2n}$, where $r$ is the normalized
strength of the perturbation and $n$ is the number of ``photons''
required for ionization.  This is consistent with conclusions drawn from
perturbation theory and other methods (the approach in \cite{[5a]} being
the closest to ours), but is, as far as we know, the first exact result
in this direction.  We also obtain, via controlled schemes, such as
continued fractions and convergent series expansions, results for strong
perturbing potentials.

Quite surprisingly our results reproduce many features of the
experimental curves for the multiphoton ionization of excited hydrogen
atoms by a microwave field \cite{[3]}.  These features include both the
general dependence of the ionization probabilities on field strength as
well as the increase in the life time of the bound state when
$-n\hbar\omega$, $n$ integer, is very close to the binding energy. Such
``resonance stabilization'' is a striking feature of the Rydberg level
ionization curves [3].  These successes and comparisons with analytical
results \cite{[1]}-\cite{[7]} suggest that the simple model we shall now
describe contains many of the essential ingredients of the ionization
process in real systems.

The model we consider is the much studied one-dimensional system with
Hamiltonian \cite{[5]}, \cite{[5a]}, \cite{[8]},

\begin{equation}
  \label{eq:(1)}
  H_0=-{\hbar^2\over 2m}{\frac{d^2}{dy^2}}-g\delta (y),\ g>0,\ \ -\infty<y<\infty.
\end{equation}
 $H_0$ has a single bound state $ u_b(y)=\sqrt{p_0}e^{-p_0|y|},\
p_0=\frac{m}{\hbar^2}g$ with energy $-\hbar \omega_0=-\hbar^2p_0^2/2m$ and a
continuous uniform spectrum on the positive real line, with  generalized
eigenfunctions 
$$u(k,y)=\frac{1}{\sqrt{2\pi}}\left
(e^{iky}-\frac{p_0}{p_0+i|k|}e^{i|ky|} \right ), \ \ -\infty<k<\infty$$

\z and energies $\hbar^2k^2/2m$.

Beginning at some initial time, say $t=0$, we apply a perturbing potential 
$ - g\eta(t) \delta(y)$, i.e.\ we change  the parameter $g$ in
$H_0$ to $g(1 + \eta(t))$ and 
solve the time
dependent Schr{\"o}dinger equation for $\psi(y,t)$,

\begin{eqnarray}
  \label{eq:(2)}
 \psi (y,t)=\theta (t)u_b(y)e^{i \omega_0 t}\hskip 4cm\nonumber\\ \hskip
1cm +\int_{-\infty}^
{\infty}\Theta (k,t)u(k,y)e^{-i\frac{\hbar k^2}{2m}t}dk \ \ (t\geq 0)
\end{eqnarray}
with initial values $\theta (0)=1,\ \Theta (k,0)=0$.  This gives the
survival probability $|\theta(t)|^2$, as well as the fraction of ejected
electrons $|\Theta(k,t)|^2 dk$ with (quasi-) momentum in the interval
$dk$.

In a previous work \cite{[8]} we found that this problem can be reduced
to the solution of a single integral equation. Using units in which
$p_0, \omega_0,\hbar,2m$ and $\frac{g}{2}$ equal $1$ we get

\begin{eqnarray}
  \label{eq:(3)}
  &\theta (t)=1+2i\int_0^t Y(s) ds \\
  &\Theta(k,t)= 2|k|/\big[\sqrt{2\pi} (1-i|k|)\big]\int_0^t Y(s) e^{i(1+k^2)s} ds
\end{eqnarray}

\z where $Y(t)$ satisfies the integral equation

\begin{equation}
  \label{eq:(5)}
  Y(t)=\eta(t)\left
\{1+\int_0^t [2i+M(t-t')]Y(t') dt'\right \}
\end{equation}

\z  with $$ M(s)=\frac{2i}{\pi}\int_0^\infty \frac{u^2e^{-is(1+u^2)}}{1+u^2}du=
\frac{1}{2}\sqrt{\frac{i}{\pi}}\int_s^\infty\frac{e^{-iu}}{u^{3/2}} du.
$$ 

An important result of the present work is that {\em when $\eta(t)$ is a
trigonometric polynomial with real coefficients
\vskip -0.4cm
\begin{equation}
  \label{eq:(7)}
 \eta(t)=\sum_{j=1}^{n}A_j\sin(j\omega
  t)+\sum_{j=1}^{m}B_j\cos(j\omega t)
\end{equation}
\vskip -0.4cm \z the survival probability $|\theta(t)|^2$ tends to zero
as $t\rightarrow\infty$, for all $\omega>0$.}

This result follows from (\ref{eq:(3)}) and
(\ref{eq:(5)}) once we establish that $2|Y(t)| = |\theta^\prime(t)| \to
0$ in an integrable way, and this represents the difficult part of
the proof. Since the main features of the behavior of $y(p)$ are already
present in the simplest case $\eta=r\sin(\omega t)$ we now specialize to this
case.  The asymptotic characterization of $Y$ is obtained from its
Laplace transform $y(p) = \int^\infty_0 e^{-pt} Y(t) dt$, which
satisfies the functional equation (cf. (\ref{eq:(5)}))
\vskip -0.2cm
\begin{eqnarray}
  \label{eq:(8)}
y(p) = \frac{ir}{2}\left\{{y(p+i\omega )\over \sqrt{1-ip+\omega} - 1} -
{y(p-i\omega) \over \sqrt{1-ip-\omega}-1}\right\}\nonumber\\+{r\omega \over
\omega^2+p^2}\ \ \ \ \ \ \ \ \ 
\end{eqnarray}
\vskip -0.2cm
\z with the boundary condition $y(p)\rightarrow 0$ as $\Im(p)\rightarrow
\pm\infty$ (the relevant branch of the square root is
$(1-ip-\omega)^{1/2} = -i(\omega-1+ip)^{1/2}$ for $\omega > 1$). We show
that the solution of (\ref{eq:(8)}) with the given boundary conditions
is unique and analytic for $\Re(p)>0$, and its only singularities on the
imaginary axis are square-root branch points (see below).  This in turn
implies that $|Y(t)|$ does indeed decay in an integrable way.  The proof
depends in a crucial way on the behavior of the solutions of the
homogeneous equation associated to (\ref{eq:(8)}): $y(p)$ has poles on a vertical
line if the homogeneous equation has a solution that is uniformly
bounded along that line. The absence of such solutions in the closed
right half plane is shown by exploiting the symmetry with respect to
complex conjugation of the underlying physical problem and carries
through directly to the more general periodic potential (6).

To understand the ionization processes as a function of $t$, $\omega$,
and $r$ requires a detailed study of the singularities of $y(p)$ in the
whole complex $p$-plane.  This yields the following results: For small
$r$, $y(p)$ has square root branch points at $p=\{-i(n\omega+1)+O(r^2):\
n\in\ZZ\}$, is analytic in the right half plane and also in an open
neighborhood ${\cal{N}}$ of the imaginary axis with cuts through the
branch points. As $|q|\rightarrow\infty$ in ${\cal{N}}$ we have
$|y(q)|=O(r \omega |q|^{-2})$.

If $|\omega-\frac{1}{n}|> {\rm const.}r^2,\,n$ a positive integer, then
for small $ r $ the function $y$ is meromorphic in the strips $
-m\omega-1-O(r^2)>\Im(p)>-m\omega-\omega-1+O(r^2),\,\ m\in\ZZ$ and has
a unique pole in each of these strips, at a point $p$ with
$0>\Re(p)=O( r ^{2n})$ for small $ r $.  {\em It then follows that
$\theta(t)$ can be decomposed as {\rm\cite{[9]}}

\begin{eqnarray}
  \label{eq:intform}
 \theta(t)=
e^{-\gamma(r;\omega) t}e^{it}F_\omega(t)+\sum_{m=-\infty}^\infty
e^{i(1+m\omega)t}h_m(t)&
\end{eqnarray}

\z where  $F_\omega$ is periodic of period
$2\pi\omega^{-1}$ and its Fourier coefficients decay faster than $r^n
n^{-n/2}$, and $|h_m(t)|\le const. r^m t^{-3/2}$ for large $t$ uniformly
in $m$. Furthermore, $h_m(t)\sim \sum_{j=0}^{\infty}c_{m,j}t^{-3/2-j}$
for large $t$.}

\begin{figure}
\epsfig{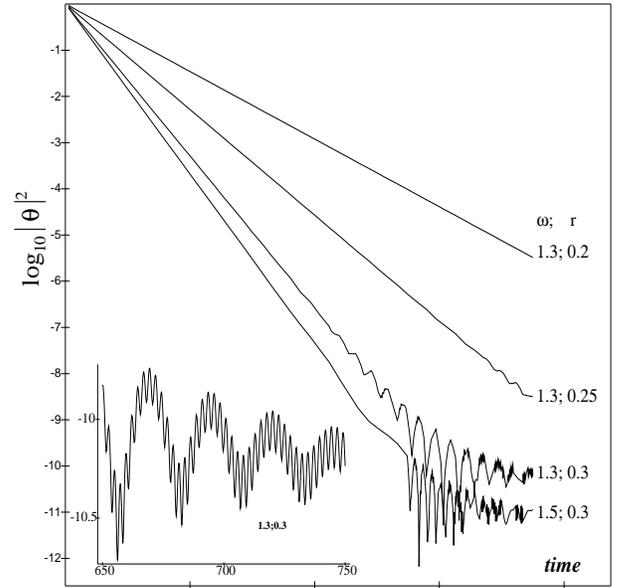}
\caption{Plot of $\log_{10}|\theta(t)|^2$ vs. time in units of
$\omega_0^{-1}$ for several values of $\omega$ and $r$. Inset shows
detail
of power-law tail for $\omega=1.3,\,r=0.3$.}
\end{figure}

Consequently, for times of order $1/\Re(\gamma)$ the survival probability
decays  as $\exp(-\Gamma t)$, $\Gamma = 2\Re(\gamma)$, after
which its long time behavior is $|\theta(t)|^2=O(t^{-3})$. This is
illustrated in Figure 1 where it is seen that for small $r$ exponential decay holds up to
times at which the survival probability is extremely small, after which
$|\theta(t)|^2$ decays polynomially with many oscillations. 
Note that even for $r$ as large as $0.3$ the decay is essentially purely
exponential for all practical purposes. Thus, for $\omega>1$ Fermi's
golden rule works magnificently \cite{[1]}.

Using a continued fraction representation of the solutions of the
homogeneous equation associated to (\ref{eq:(8)}) we obtain as
$r\rightarrow 0$,
\begin{equation}\label{devilstaircase}
\Gamma=\left\{\begin{array}{lllllll} \displaystyle
      \sqrt{\omega-1}\frac{ r ^2}{\omega}; &\mbox{\rm if }{\omega} >1+O(r^2)
      \\ &\\
  \displaystyle 
\frac{ \sqrt{2\omega -1}}{(1-\sqrt{1-\omega})^2}\frac{ r
  ^4}{8\omega};&\mbox{\rm if }{\omega} \in (\frac{1}{2},1)^+
\\
\ldots &\ldots
\\ \displaystyle\frac{2^{-2n+2}\sqrt{n\omega-1}}{\prod_{m<
n}(1-\sqrt{1-m\omega})^2}\frac{ r ^{2n}}{n\omega}; &\mbox{\rm if }{\omega}\in
(\frac{1}{n},\frac{1}{n-1})^+
\end{array}\right.\end{equation}\vskip -0.2cm 
\z where $\omega\in (a,b)^+$ means $a+O(r^2)<\omega<b-O(r^2)$. The
result for $\omega>1$ agrees with perturbation theory \cite{[1]} since the
the transition matrix element is
\vskip -1cm
\begin{equation}\label{e10}
\big|<u_b(y)|\delta(y)u(k,y)>\big|^2=\frac{1}{2\pi}\frac{k^2}{1+k^2}.
\end{equation}
\z
In Figure 2 we plot the behavior of $\Gamma^{-1}$ which is just the time
needed for $|\theta(t)|^2$ to decay significantly, as a function of
$\omega$.  
\begin{figure}
\epsfig{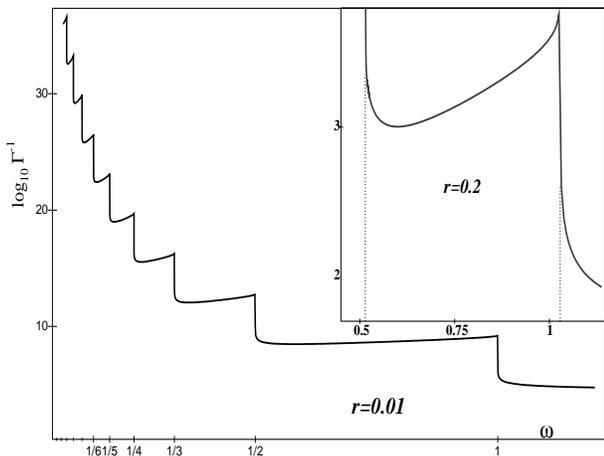}
\caption{$\log_{10}\Gamma^{-1}$ as a function of $\omega/\omega_0$ at
$r=0.01$. At $r=0.2$ (inset)  shift of the resonance is visible. }
\end{figure}
The curve is made up of smooth (roughly self-similar) pieces for
$\omega$ in the intervals $(n^{-1}, (n-1)^{-1})$ corresponding to
ionization by $n$ photons.  Note that at resonances, when $\omega^{-1}$
is an integer (i.e.\ multiple of $\omega_0^{-1}$ here set equal to
unity), the coefficient of $r^{2n}$, the leading term in $\Gamma$, goes
to zero.  At such values of $\omega$ one has to go to higher order in
$r$, corresponding to letting $\omega$ approach the resonance from
below.  This yields an enhanced stability of the bound state against
ionization by perturbations with such frequencies. The origin of this
behavior is, in our model, the vanishing of the matrix element in
(\ref{e10}) at $k=0$. This behavior should hold quite generally since the
quasi-free wavefunction $u(k,y)$ may be expected to vanish pointwise as
$k\rightarrow 0$. For $d\ge 1$ there is an additional factor $k^{d-2}$
coming from the energy density of states near $k=0$.  As $r$ increases
these resonances shift in the direction of increased frequency.  For
small $r$ and $\omega=1$ the shift in the position of the resonance,
sometimes called the dynamic Stark effect [1], is about
$\frac{r^2}{\sqrt{2}}$.

In Figure 3 we plot the strength of the perturbation $r$, required to
make $|\theta(t)|^2 = {1 \over 2}$ for a fixed number of oscillations of
the perturbing field (time measured in units of $\omega^{-1}$) as a
function of $\omega$.  Also included in this figure are experimental
results for the ionization of a hydrogen atom by a microwave field.  In
these still ongoing beautiful series of experiments, carried out by
several groups and reviewed in \cite{[3]}, the atom is initially in an
excited state with principal quantum number $n_0$ ranging from 32 to
90. The experimental results in Fig. 3 are taken from Table 1 in
\cite{[3]}, see also Figures 13 and 18 there.  The ``natural frequency''
$\omega_0$ is there taken to be that of a transition from $n_0$ to $n_0
+1$, $\omega_0 \sim n^{-3}_0$.  The strength of the microwave field $F$
is then normalized to the strength of the nuclear field in the initial
state, which scales like $n^{-4}_0$.  The plot there is thus of $n^4_0
F$ vs. $n^3_0 \omega$.  To compare the results of our model with the
experimental ones we had to relate $r$ to $n_0^4 F$. Given the
difference between the hydrogen atom Hamiltonian with potential
$V_0(R)=-1/R$ perturbed by a polarized electric field $V_1=xF\sin(\omega
t)$, and our model with $V_1=rV_0$, this is clearly not something that
can be done in any unique way.  We therefore simply tried to find a
correspondence between $n_0^4F$ and $r$ which would give the best visual
fit. Somewhat to our surprise these fits for different values of
$\omega/\omega_0$ all turned out to have values of $r$ close to
$3n_0^4F$. A correspondence of the same order of magnitude is obtained
by comparing the perturbation-induced shifts of  bound state energies in
our model and in Hydrogen.

\begin{figure}
\epsfig{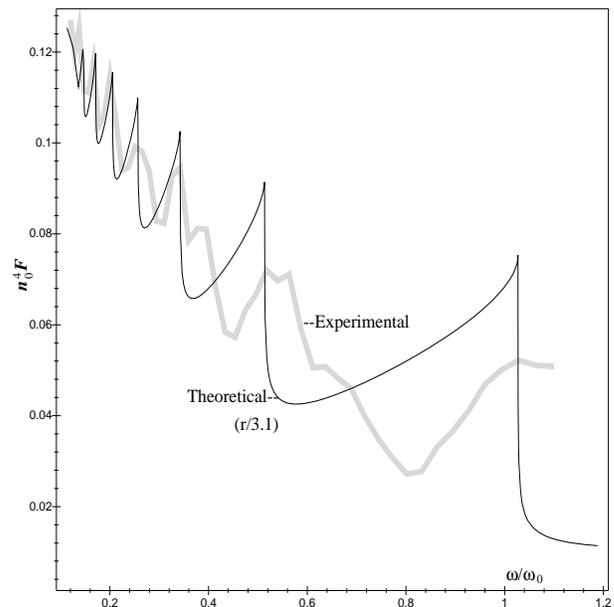}
\caption{Comparison of the theoretical and experimental threshold
amplitudes for $50\%$ ionization vs. $\omega/\omega_0$.}
\end{figure}
The shift in the position of the resonances from the integer fractional
values seen in Fig. 2, due to the finite value of $r$, was approximated
in Fig. 3 using the average value of $r$ over the range, $r\approx
0.195$.

In Figure 4 we plot $|\theta(t)|^2$ vs. $r$ for a fixed $t$ and two
different values of $\omega$. These frequencies are chosen to correspond
to the values of $\omega/\omega_0$ in the experimental curves.  Figure 1
in \cite{[10]} and Figure 1b in \cite{[3]}. The agreement is very good
for $\omega/\omega_0\approx .1116$ and reasonable for the larger ratio.
Our model essentially predicts that when the fields are not too strong,
the experimental survival curves for a fixed $n_0^3 \omega$ (away from
the resonances) should behave essentially like
$\displaystyle\exp\left(-C[n^4_0 F]^{\scriptstyle\frac{\scriptstyle
2}{\scriptstyle n_0^{3}\omega}}\,\,t\omega\right)$ with $C$ depending on
$n_0^3\omega$ but, to first approximation, independent of $n_0^4 F$.

The degree of agreement between the behavior of what might be considered
as the absolutely simplest quantum mechanical model of a bound state
coupled to the continuum and experiments on hydrogen atoms is truly
surprising.  The experimental results and in particular the resonances
have often been interpreted in terms of classical phase space orbits in
which resonance stabilization is due to KAM--like stability islands
\cite{[3]}.  Such classical analogs are  absent in our model
as in fact are ``photons''. On the other hand, the special nature of the
edge of the continuum seems to be quite general, cf. \cite{[5a]}.

\begin{figure}
\epsfig{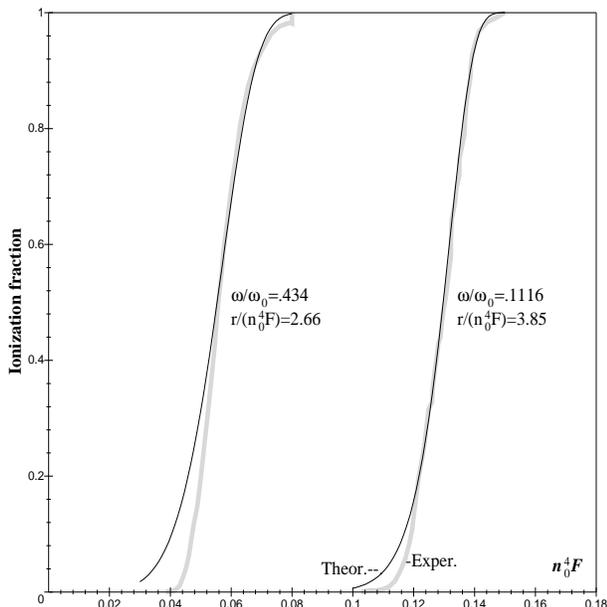}
\caption{Ionization fraction at fixed $t$ (corresponding to 300
oscillations) as a function of amplitude.}
\end{figure}

We note that for $\omega>\omega_0$, in the limit of small amplitudes
$r$, a predominantly exponential decay of the survival probability
followed by a power-law decay was proven in \cite{[6]} for three
dimensional models with quite general local binding potentials having one
bound state, perturbed by a local potential of the form $r\cos(\omega
t)V_1(y)$. It seems likely that our results for general $\omega$ and
$r$, including general periodic (perhaps also quasi-periodic)
perturbations would extend to a similarly general setting. We are
currently investigating various extensions of our model to understand
the effect of the restriction to one bound state. This will hopefully
lead to a more detailed understanding, and some control over the
ionization process.

Because $\Gamma$ relates to the position of the poles of the solution of
(\ref{eq:(8)}), a convenient way to determine $\Gamma$ (mathematical
rigor aside), if $r$ is not too large, is the following, see also
\cite{[5a]}.  One iterates $n$ times the functional equation
(\ref{eq:(8)}), $n$ appropriately large, to express $y(p)$ only in terms
of $y(p\pm mi\omega)$ with $|m|> n$. After neglecting the small
contributions of the $y(p\pm mi\omega)$, the poles of $y(p)$ can be
obtained by a rapidly converging power series in $r$, whose coefficients
are relatively easy to find using a symbolic language program, although
a careful monitoring of the square-root branches is required.
A complete  study of the poles and branch-points of $y$ leads to
(\ref{eq:intform}) which is effectively the Borel summation of the
formal (exponential) asymptotic expansion of $Y$ for
$t\rightarrow\infty$.

\z Acknowledgments.  We thank A. Soffer, M. Weinstein and P. M. Koch for
valuable discussions and for providing us with their papers. We also
thank R. Barker, S. Guerin and H. Jauslin for introducing us to the
subject.  Work of O. C. was supported by NSF Grant 9704968, that of J.
L. L. and A. R. by AFOSR Grant F49620-98-1-0207.

\smallskip

\z * Also Department of Physics.

\z costin\symbol{64}math.rutgers.edu, lebowitz@sakharov.rutgers.edu,
rokhlenk@math.rutgers.edu.

\end{document}